\documentclass[prl,showpacs,twocolumn,amsmath,amssymb]{revtex4}

\usepackage{bm}
\usepackage{graphicx}
\usepackage{bbold}

\newcommand{\tth}{$^{\text{th}}$\ }
\newcommand{\pdagger}{{\phantom{\dagger}}}
\newcommand{\dt}{\Delta\tau}
\newcommand{\PT}{_{\text{PT}}}
\newcommand{\refq}[1]{(\ref{eq:#1})}
\newcommand{\reff}[1]{Fig.~\ref{fig:#1}}

\usepackage{color}

\definecolor{darkgreen}{rgb}{0,0.7,0}

\begin{document}

\title{The Mott insulator -- 10\tth order perturbation theory\\ 
extended to infinite order using QMC}

\author{N.~Bl\"umer}
\email{Nils.Bluemer@uni-mainz.de}
\affiliation{Institute of Physics, Johannes Gutenberg University, 55099 Mainz, Germany}
\author{E. Kalinowski}
\affiliation{Department of Physics, Philipps University, 35032 Marburg, Germany}

\date{\today}

  \begin{abstract}
    We present a new method, based on the combination of analytical
    and numerical techniques within the framework of the dynamical
    mean-field theory (DMFT). Building upon numerically exact results
    obtained in an improved quantum Monte Carlo (QMC) scheme, 10\tth
    order strong-coupling perturbation theory for the Hubbard model on
    the Bethe lattice is extrapolated to infinite order. We obtain
    continuous estimates of energy $E$ and double occupancy $D$ with
    unprecedented precision ${\cal O}(10^{-5})$ for the Mott insulator
    above its stability edge $U_{c1}\approx 4.78$ as well as critical
    exponents.
    The relevance for recent experiments on Cr-doped V$_2$O$_3$ is pointed out.
  \end{abstract}
  \maketitle

  The exploration of the transition from a paramagnetic metallic
  phase to a paramagnetic Mott-Hubbard insulator is of fundamental
  interest in condensed matter theory
  \cite{multi:MIT}.  
  Essential features of this correlation phenomenon are
  captured by the single-band Hubbard model
  \begin{equation}\label{eq:HM}
    \hat{H} = -\sum_{(ij),\sigma} t_{ij}
             \left(\hat{c}_{i\sigma}^\dagger \hat{c}_{j\sigma}^\pdagger
             + \hat{c}_{j\sigma}^\dagger \hat{c}_{i\sigma}^\pdagger\right) 
             + U \hat{D}\,,
  \end{equation}

\vspace{-0.5em}\noindent
  where the operator $\hat{c}_{i\sigma}^\dagger$
  ($\hat{c}_{i\sigma}^\pdagger$) creates (annihilates) an electron
  with spin $\sigma\in\{\downarrow,\uparrow\}$ on site $i$ and $\hat
  D=\sum_{i}\hat n_{i\uparrow}\hat n_{i\downarrow}$ 
counts the number of doubly occupied sites;
  $\hat{n}_{i\sigma}=\hat{c}_{i\sigma}^\dagger\hat{c}_{i\sigma}^\pdagger$.
  Its nonperturbative solution in the thermodynamic limit is possible
  within the framework of
  the dynamical mean-field theory (DMFT) which maps the lattice problem
  onto an Anderson impurity model; this mapping becomes exact in
  the limit of infinite dimensionality  \cite{multi:DMFT}.
  The nature of the
  Mott-Hubbard transition in the idealized case of a fully frustrated Hubbard 
  model with semi-elliptic \cite{note:DOS} noninteracting density of states 
  (DOS) has been a matter of intense debate
  \cite{multi:MIT_Bethe,Noack99a}.
  As a result, a coexistence region of metallic and insulating solutions,
  enclosed by spinodal lines $U_{c1}(T)<U_{c2}(T)$,
  has been established below a critical temperature $T^*\approx 0.06t$. 
  Only recently, 
  estimates of energy $E=\langle \hat{H}\rangle/N$ and
  double occupancy $D=\langle \hat{D}\rangle/N$ per lattice site
  have been obtained in quantum Monte Carlo (QMC) calculations with
  high enough precision for reliably 
  pinpointing the first-order metal-insulator transition (MIT) line 
  \cite{Bluemer02a}. 
  A complete and universally accepted \cite{note:non_first}
  understanding of this prototype MIT is highly desirable (i) as a
  prerequisite for controlled studies of more material-specific
  aspects of such transitions (e.g., band structure, degree of 
  frustration)  and (ii) since already the single-band description can
  explain surprisingly many aspects of real-world experiments, which
  suggests a high degree of universality. In fact, recent experiments
  on V$_2$O$_3$ have not only found agreement with critical exponents
  at $T^*$ predicted by single-band DMFT
  \cite{Limelette2003,Kotliar2003}, but even observed well-defined
  spinodal lines. The interesting question to what extent these lines
  can be identified with those computed in purely electronic models is
  still open; one criterion will be determined in this Letter.

  
  In this work, we present results for
  the ground state insulating phase (``Mott insulator'') with
  unprecedented precision. 
  After introducing the Kato-Takahashi strong coupling perturbation
  theory (PT) used for an exact calculation of all contributions to the 
  energy including order $t^{10}/U^9$, we will sketch our advanced DMFT-QMC 
  scheme which permits
  precisions of ${\cal O}(10^{-5})$ (in $E/t$ or $D$) {\em at selected
  phase points}.
  The error bounds will be confirmed at
  moderately strong coupling ($U=6t$). Finally, the breakdown of 
  finite-order PT observed at smaller $U$ will be overcome using
  an extrapolation scheme to infinite order (``ePT'') 
  which is controlled by its excellent agreement with QMC 
  throughout the coexistence region. In addition to an estimate for 
  $U_{c1}$ and to
  continuous estimates for $E$ and $D$ {\em at all coupling strengths} 
  $U\ge U_{c1}$, we will extract critical exponents from ePT (which are 
  clearly out of reach for both QMC and plain PT). 
  Beyond the direct physical interest, our results will provide 
  an essential reference in the (now very 
  active) development of new methods for solving (multi-band) DMFT, its cluster
  extensions, and hybrid schemes such as LDA+DMFT.
  The infinite-order extrapolation method ePT constructed in this Letter
  is similar in spirit to Singh's vastly successful extrapolation of 10\tth
  order PT in $J_\perp/J_\|$ for the $2d$ Heisenberg antiferromagnet
  to the isotropic limit \cite{Singh89a}. The common general ideas 
  appear promising for many areas of physics, potentially including QCD.

  {\it Perturbation theory --} 
  At half filling and in the absence of hopping, i.e. for $t_{ij}\equiv 0$ 
  in \refq{HM},
  all states without double occupancies have the minimal energy
  $\langle U \hat D\rangle=0$; due to the arbitrariness of the spin
  direction on each site, the degeneracy of the space $\mathcal{U}_0$
  of unperturbed ground states is $2^N$.  An expansion around this
  strong-coupling limit at $T=0$ using Kato's formalism \cite{Kato49}
  was established by Takahashi \cite{Takahashi77a}; for details in the
  case at hand, see Refs.~\onlinecite{Eastwood03} and 
  \onlinecite{Kalinowski02}.  In this
  formalism, every ground state $\psi_n$ of $\hat H$ is
  obtained by application of a well-defined operator $\hat \Gamma$ on
  a state $\phi_n \in \mathcal{U}_0$: $\psi_n=\hat\Gamma \phi_n$.
  Norm conservation implies
  $\hat\Gamma^{\dagger}\hat\Gamma=\hat P_0$, where $\hat P_0$ is the
  projector on $\mathcal{U}_0$. Thus, the Schr\"odinger
  equation  $\hat H \psi_n=E_0 \psi_n$ may be written as
  \[
    \hat h\,\phi_n=E_0\,\phi_n\,\text{,}\quad\quad 
    \hat h\equiv\hat\Gamma^{\dagger}\hat H\hat\Gamma\,\text{,}
  \]
  where $\hat h$ acts as an effective operator in the unperturbed
  space and can be expanded in orders of $t/U$:
  \[
    \hat h=U \sum\nolimits_{m=1}^{\infty}\hat h_m\,\left(t/U\right)^m\,\text{.}
  \]
  The operators $\hat h_m$ generate electron transfers; since they
  have to preserve the absence of double occupancies, the associated
  diagrams are always closed. Specifically for the Bethe lattice, all
  closed paths are self-retracing. Neither unlinked
  diagrams nor diagrams of odd order in $t$ contribute (at half filling)
  for any lattice
  type \cite{Takahashi77a}. Finally, graphs with more than two lines
  between any pair of sites are suppressed at least as $1/Z$.
  \reff{PT_diag} illustrates the
  first three contributing orders. For the half-filled Hubbard model,
  $m$\tth order Kato-Takahashi PT yields coefficients
  for a generalized Heisenberg model including up to $m$-site
  interactions.  In the paramagnetic phase and for $Z=\infty$, an
  exact evaluation is possible since then all spin-spin correlation
  functions vanish.
  \begin{figure}
  \includegraphics[width=\columnwidth,clip=true]{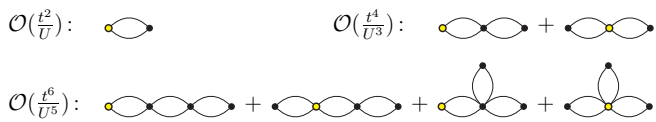}
  \caption{Low order diagrams in PT: 
    A first hopping process creates an empty site (circle). After a 
    total of $2n$ hoppings, impacting $n$ other
    sites (spheres), the ground state is restored.}
  \label{fig:PT_diag}
  \end{figure}
  The 10\tth order result \refq{E_PT} obtained from a 
  computer implementation \cite{Kalinowski04} of this diagrammatic approach 
  will be discussed below.

  {\it Quantum Monte Carlo --}
  Deep enough within the insulating phase, all excitations of the
  insulator are exponentially suppressed. Consequently, ground state
  properties can be estimated from measurements at finite (but small)
  temperatures $T$ (and vice versa); the precision can be tested and further increased
  in explicit extrapolations $T\to 0$. At finite $T$, the
  impurity part of the DMFT self-consistency problem may be solved
  using the auxiliary field quantum Monte Carlo (QMC)
  method~\cite{Hirsch86a}; its accuracy is primarily limited by the
  discretization $\dt$ of the imaginary time $0\le \tau\le T^{-1}$. In
  general, the $\dt$ error consists of the Trotter error and an error
  associated with discrete Fourier transformations (FT) performed
  twice per iteration of the DMFT self-consistency cycle. The FT error
  is essentially eliminated in a new high-precision QMC algorithm
  where the low-frequency physics of the impurity is obtained from QMC
  while the high-frequency part is computed
  analytically. This method converges much faster for $\dt\to 0$ at all
  frequencies than previously used methods~\cite{Knecht02a}. The QMC
  results shown below are based on simulations at $T=1/15$
  for discretizations $0.09\le \dt\le 0.25$ (with quartic extrapolation
  $\dt\to 0$) using some 40
  iterations with ${\cal O}(10^7)$ sweeps each.  
  For $U\le5.0$,
  finite-$T$ corrections were estimated using additional simulations
  at $T=1/20$ and $T=1/25$ \cite{note:ins_stable}.  

  {\it Results I --}
  In the following, results are presented for the ground state (lower
  index 0 omitted).  PT for the half-filled Hubbard model \refq{HM}
  with Bethe DOS ($W=4t$) yields the energy $E(U)=E\PT(U)+ {\cal
    O}(t^{12}\, U^{-11})$, where
  \begin{equation}\label{eq:E_PT}
    E\PT(U)=-\frac{1}{2\, U} -\frac{1}{2\,U^3} -\frac{19}{8\, U^5} 
    -\frac{593}{32\, U^7} -\frac{23877}{128\, U^9}
  \end{equation}
  for $t=1$.
  Consequently, the double occupancy $D(U)=dE(U)/dU$ reads
  $D(U)=D\PT(U)+ {\cal O}(t^{12}\, U^{-12})$, i.e.,
  \begin{equation}\label{eq:D_PT}
    D\PT(U)=\frac{1}{2\, U^2} +\frac{3}{2\,U^4} 
    +\frac{95}{8\, U^6} +\frac{4151}{32\, U^8} +\frac{214893}{128\, U^{10}}\,.
  \end{equation}
  These PT results for $E$ and $D$ are shown as solid lines in the
  left and right panels of \reff{PT_QMC}, respectively.
  \begin{figure}
  \includegraphics[width=\columnwidth,clip=true]{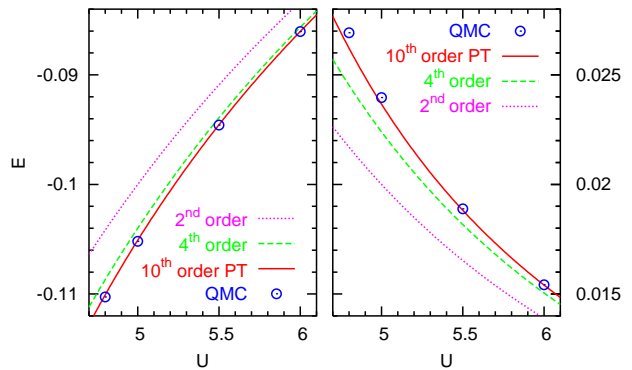}
  \caption{Ground state energy $E$ and double occupancy $D$: PT according to 
    \refq{E_PT} and \refq{D_PT} (solid lines) in comparison with
    low-order PT (dashed/dotted lines) and QMC (circles).}
  \label{fig:PT_QMC}
  \end{figure}
  The comparisons with $2^{\text{nd}}$ and 4\tth order PT
  (dashed/dotted lines) hint at a relatively fast convergence. In
  fact, a detailed analysis shows that the PT results at $U=6$ (just
  beyond the coexistence region) are converged within about $10^{-5}$
  for both $E$ and $D$. The agreement with QMC for this
  interaction confirms the correctness of the PT coefficients in
  \refq{E_PT} and of the QMC results within error bars.  
  At smaller $U$, however, PT deviates markedly [by ${\cal O}(10^{-3})$
  for $D$ at $U=4.8$] from the numerically exact QMC.
  In the remainder of this Letter, we will construct a new 
  method, the extended perturbation theory (ePT), which yields continuous 
  and thermodynamically consistent \cite{note:consist}
  estimates for $E$ and $D$ with even higher precision than QMC.
  In addition, we will obtain accurate estimates of the stability edge 
  $U_{c1}$ and of critical exponents.


  {\it Extended perturbation theory (ePT) --}
  The idea of the ePT is an extrapolation of the coefficients $a_n$ in
  the power series for the energy (for $t=1$), 
  \begin{equation}\label{eq:series}
    E(U) = \sum_{i=1}^\infty a_{2i}\, U^{1-2i}\,,
  \end{equation}
  from a small number of known coefficients $a_2, \cdots, a_{n_0}$ to
  infinite order.  Specializing on the case where all $a_{2i}$ have
  the same sign \cite{note:ePT_general}, we define the sequence
  \begin{equation}\label{eq:Uc1_def}
  U_{c1}[2i+2]\equiv\sqrt{a_{2i+2}/a_{2i}}\,.
  \end{equation}
  If this sequence converges, the ratio criterion implies the
  convergence of series \refq{series} for $U>U_{c1}\equiv\lim_{i\to\infty}
  U_{c1}[2i]$ and divergence below. More specificly, we assume
  \begin{equation}\label{eq:fit}
    U_{c1}[n] = U_{c1}+u_1/n + u_2/n^2 + {\cal O}(n^{-3})\,.
  \end{equation}
  Our assumption of this functional form will be strongly supported
  by fits to the known PT coefficients (cf.~\reff{PT_ext}). By virtue
  of such a fit, series \refq{series} and its derivatives may be evaluated at
  arbitrary order. Comparisons of the numerical infinite-order
  limits of this scheme for $E$ and $D$ with QMC 
  (cf.~\reff{dE10} and \reff{dD10})
  will then give ``numerical proofs'' of the validity of \refq{fit}.
  Let us, however, first analyze the critical behavior: Eq.~\refq{fit} 
  and \refq{Uc1_def} imply
  $\alpha_n\equiv a_n U_{c1}^{-n}= n^{-\tau} [c+{\cal O}(1/n)]$, where
   $\tau=-u_1/U_{c1}$.  Re-expanding \refq{series} around the critical
  point $U_{c1}$, we obtain
  \begin{eqnarray}
  E(U)\! &=& U_{c1} \sum\nolimits_{i=1}^\infty \alpha_{2i}\, (1+x)^{1-2i}\,;\quad
  x \equiv \frac{U-U_{c1}}{U_{c1}}\,,\nonumber\\\label{eq:Ecritical}
  &=& U_{c1} \big[\big(\sum_{n=0}^\infty e_n x^n\big)
      + e_{\tau-1}x^{\tau-1} + e_{\tau} x^{\tau} + \dots\big]\,.\quad\,
  \end{eqnarray}
  Note that any finite order in \refq{series} generates terms to all
  integer orders in $x$ in \refq{Ecritical}. The leading singular
  contribution to $E$ is associated with $e_{\tau-1}=\frac{c}{2}
  \Gamma(1-\tau)$.
  For the numerical step within the ePT we use weighted
  \cite{note:weights} quadratic least squares fits to \refq{fit} with either
  no adjustable parameters (``unrestricted'' fit) or with minimal tuning of
  $U_{c1}$ for enforcing analytic constraints or for obtaining
  error bounds from comparison with QMC.
  As seen in \reff{PT_ext},
  \begin{figure}
  \includegraphics[width=\columnwidth,clip=true]{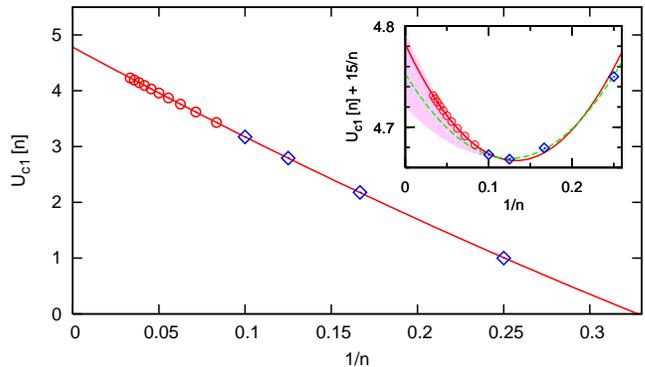}
  \caption{Construction of ePT: PT values for $U_c[n]$ 
    (squares) are extrapolated to $1/n\to 0$ using a quadratic
    fit (solid line) restricted to $\tau=3.5$. Values at
    smaller $1/n$ (circles) define ePT coefficients to all orders.
    Inset: restricted and unrestricted fit (solid/dashed line), 
    PT ``data'', and range compatible with QMC (shaded) after
    subtraction of linear term.}
  \label{fig:PT_ext}
  \end{figure}
  the known coefficients (squares) are well represented by a
  ``restricted'' fit (solid line) to be discussed below. Its
  small curvature as well as slight deviations from the PT values become
  apparent only in the inset, where an average linear term has been
  subtracted.  Also shown in the inset is the unrestricted fit (dashed line) 
  which predicts $U_{c1}=4.75$ and $\tau=3.44$. The shaded region marks
  extrapolation curves which lead to estimates for $E$ and $D$ compatible
  with QMC; thus, QMC implies error bounds $4.72\le U_{c1}\le 4.79$ and 
  $3.36\le \tau\le 3.53$. 
  Half-integer exponents occur in the related Falicov-Kimball model 
  \cite{vanDongen92a}
  and are generally expected in the context of mean-field theories.  
  For our final estimates, we therefore {\em assume} an exponent 
  $\tau=3.5$ in a restricted fit
  which implies $U_{c1}=4.782$ (in agreement with NRG 
  \cite{Bulla01a} and DMRG \cite{Garcia04a})
  and corresponds to the solid lines in \reff{PT_ext} \cite{note:12th}.
  
  {\it Results II --}
  The ePT estimates for energy $E$ and double occupancy $D$
  are shown in \reff{dE10} 
  \begin{figure}
    \includegraphics[width=\columnwidth,clip=true]{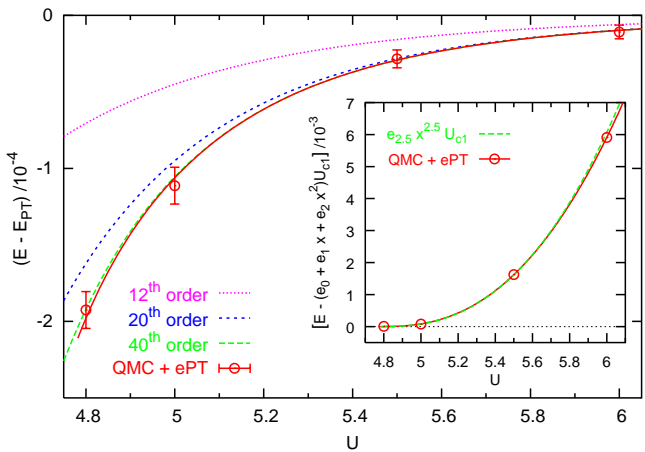}
  \caption{Main panel: Energy from QMC (circles), ePT (solid
    line), and truncated ePT (dashed/dotted lines); in
    all cases, the PT result [Eq.\ \refq{E_PT} and solid line in
    \reff{PT_QMC}] has been subtracted. Inset: energy correction in
    ePT beyond non-critical terms (solid line) in comparison with
    the leading critical term.}
  \label{fig:dE10}
  \end{figure}
  \begin{figure}
    \includegraphics[width=\columnwidth,clip=true]{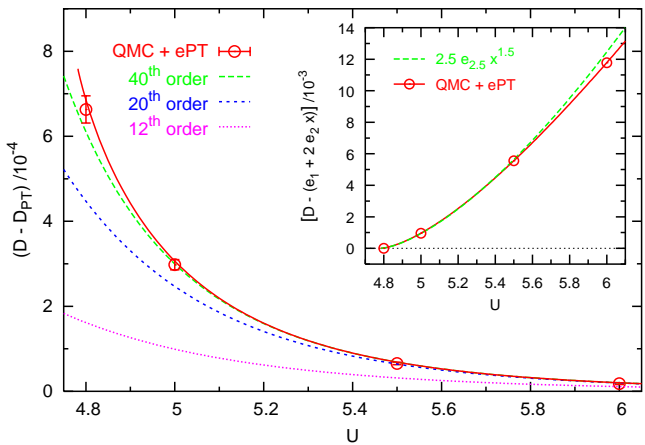}
  \caption{Double occupancy from QMC (circles) and ePT (solid lines) in 
    complete analogy with \reff{dE10}.}
  \label{fig:dD10}
  \end{figure}
  and \reff{dD10}, respectively (solid lines);
  in the main panels, the 10\tth order PT results have been
  subtracted for enhanced
  clarity. These results agree perfectly with QMC (circles) which confirms
  the thermodynamic consistency of the latter \cite{note:consist}. In
  contrast, low order truncated ePT (dotted/dashed lines) deviates
  markedly, in particular for $D$. The insets of \reff{dE10} and
  \reff{dD10} show contributions arising from the leading singular
  term in \refq{Ecritical} using $e_{2.5}=-c\frac{4}{15}\sqrt{\pi}$,
  where $c\approx -0.00725$ is derived from the asymptotics of
  $n^{3.5} \alpha[n]$. These
  semi-analytic results agree well with the QMC/ePT estimates for contributions
  including and beyond the leading critical term. Thus,
  the critical behavior is confirmed {\em a posteriori} by
  QMC. The polynomial (in $x$) representation \cite{note:fit}
  of the ePT results used in the above analysis
  is more accurate than 10\tth order PT for $U< 6.6$ and 
  should be used in future comparisons.
  
  {\it Summary --}
  We have combined the strengths of high-order PT and of
  an improved QMC scheme and have constructed the ePT, an extension of PT to
  infinite order, in order to compute ground state properties
  of the Mott insulator within DMFT with unprecedented accuracy.
  We have confirmed earlier and independent estimates
  \cite{Bulla01a,Garcia04a} for $U_{c1}$, the lower (meta)stability
  edge of the insulator, and for the first time determined critical
  exponents ($5/2$ for $E$, $3/2$ for $D$). 
  As both the spinodal line $U_{c1}(T)$ and the thermodynamic
  observables within the insulating phase are nearly $T$-independent
  (for $T\lesssim T^*/2$),
  these results are also excellent approximations at finite
  temperatures  where
  the MIT and both spinodal lines have been observed experimentally
  in Cr doped V$_2$O$_3$
  \cite{Limelette2003,Kotliar2003}. In particular, a comparison
  between experimental estimates for the critical exponents and our
  predictions could shed light on the still ill-understood role of the
  lattice at the transition.
  With the authoritative results for $E$ and $D$ provided by ePT and
  QMC \cite{note:Pozgajcic2004}, all details of the first-order Mott
  transition can now be clearly resolved which eliminates any room for
  alternative continuous-transition scenarios \cite{note:non_first}. 
  Complementary results for the metallic ground state can be found in Ref.\ 
  \onlinecite{Bluemer04b}; various finite-temperature results 
  will be published elsewhere.  
  
  We thank F.~Gebhard for stimulating our interest in the subject,
  P.G.J.~van Dongen for help in determining the critical exponents,
  and both as well as W.~Gluza, K.~Held, E.~Jeckelmann, and C.~Knecht
  for discussions.\vspace{-0.3cm}

  \bibliography{mottins}

\begin{thebibliography}{28}
\bibitem[{not({\natexlab{a}})}]{note:DOS}
\bibinfo{howpublished}{This DOS is realized on a Bethe lattice with
  nearest-neighbor hopping $t_{ij}=t/\sqrt{Z}$ for coordination $Z\to\infty$.}
\bibitem[{not({\natexlab{c}})}]{note:ins_stable}
\bibinfo{howpublished}{Note that the insulating phase is thermodynamically
  stable at all phase points evaluated within QMC.}
\bibitem[{not({\natexlab{d}})}]{note:consist}
\bibinfo{howpublished}{Note that within QMC $E$ and $D$ are essentially
  independent while $D=dE/dU$ within (e)PT by construction.}
\bibitem[{not({\natexlab{e}})}]{note:ePT_general}
\bibinfo{howpublished}{Oscillations observed in PT for finite $d$ would
  require, e.g., individual ePT fits for subsets of coefficients.}
\bibitem[{not({\natexlab{h}})}]{note:fit}
\bibinfo{howpublished}{The coefficients in \refq{Ecritical} are in terms of
  $\tilde e_n\equiv U_{c1} e_n$ (for $U_{c1}=4.782$): $\tilde e_0=-0.1107586$,
  $\tilde e_1=0.1302062$, $\tilde e_2=-0.2213925$, $\tilde e_{2.5}=0.18536$,
  $\tilde e_3=0.04532$, $\tilde e_{3.5}=-0.13$, $\tilde e_4=0.03406$, $\tilde
  e_5=0.03759$, $\tilde e_6=-0.03$.}
\end{thebibliography}
 \end{document}